\documentstyle [eqsecnum,preprint,aps,epsf]{revtex}

\makeatletter
\def\footnotesize{\@setsize\footnotesize{10.0pt}\xpt\@xpt
\abovedisplayskip 10\p@ plus2\p@ minus5\p@
\belowdisplayskip \abovedisplayskip
\abovedisplayshortskip  \z@ plus3\p@
\belowdisplayshortskip  6\p@ plus3\p@ minus3\p@
\def\@listi{\leftmargin\leftmargini
\topsep 6\p@ plus2\p@ minus2\p@\parsep 3\p@ plus2\p@ minus\p@
\itemsep \parsep}}
\long\def\@makefntext#1{\parindent 5pt\hsize\columnwidth\parskip0pt\relax
\def\strut{\vrule width0pt height0pt depth1.75pt\relax}%
$\m@th^{\@thefnmark}$#1}
\long\def\@makecaption#1#2{%
\setbox\@testboxa\hbox{\outertabfalse %
\reset@font\footnotesize\rm#1\penalty10000\hskip.5em plus.2em\ignorespaces#2}%
\setbox\@testboxb\vbox{\hsize\@capwidth
\ifdim\wd\@testboxa<\hsize %
\hbox to\hsize{\hfil\box\@testboxa\hfil}%
\else %
\footnotesize
\parindent \ifpreprintsty 1.5em \else 1em \fi
\unhbox\@testboxa\par
\fi
}%
\box\@testboxb
} %
\global\firstfigfalse
\global\firsttabfalse
\def\tabular{\let\@halignto\@empty\@tabular}
\def\endtabular{\crcr\egroup\egroup $\egroup}
\expandafter \def\csname tabular*\endcsname #1{\def\@halignto{to#1}\@tabular}
\expandafter \let \csname endtabular*\endcsname = \endtabular
\def\@tabular{\leavevmode \hbox \bgroup $\let\@acol\@tabacol
   \let\@classz\@tabclassz
   \let\@classiv\@tabclassiv \let\\\@tabularcr\@tabarray}
\def\endtable{%
\global\tableonfalse\global\outertabfalse
{\let\protect\relax\small\vskip2pt\@tablenotes\par}\xdef\@tablenotes{}%
\egroup
}%
\global\firstfigfalse
\global\firsttabfalse

\mathcode`\:="003A%
\mathchardef\:="303A%
\mathchardef\lt="313C%
\mathchardef\gt="313E%
\mathcode`\@="8000 {\catcode`\@=\active \gdef@{{\dagger}}}%
\mathcode`\<="8000 {\catcode`\<=\active \gdef<{{\delimiter"426830A}}}%
\mathcode`\>="8000 {\catcode`\>=\active \gdef>{{\delimiter"526930B}}}%
\def\undertilde{\mathpalette\utild@}%
\newbox\utildbox
\def\utild@#1#2%
    {%
    \setbox0=\hbox {$\m@th #1 {#2}$}%
    \dimen@=\dp0\advance\dimen@ by 0.2 ex\dp0=\dimen@%
    \setbox\utildbox=\hbox {$\tilde {\hphantom {\copy0}}$}%
    \setbox0=\vtop {\offinterlineskip\box0\box\utildbox}%
    \dp0=0.4ex\box0\relax%
    }%
\let\u=\undertilde
\makeatother

\thicklines

\def\triad#1#2{
   \put(0,0){\vector(1,0){5}}
   \put(0,0){\vector(0,1){5}}
   \put(0,0){\vector(-1,-1){3}}
   \put(0,7){\makebox(0,0){#1}}
   \put(7,0){\makebox(0,0){#2}}
   \put(-3,-5){\makebox(0,0){$x_\perp$}}
}

\def\FigSix{
   \begin{picture}(100,60)(-65,-35)
   \put(-50,-20){
      \begin{picture}(50,40)(-25,-20)
      \put(-25,-10){\line(1,0){50}}
      \put(-25, 10){\line(1,0){50}}
      \triad{$\tau$}{$z$}
      \put(  0, 12){\makebox(0,0){$\tau=\beta$}}
      \put(  0,-12){\makebox(0,0){$\tau=0$}}
      \end{picture}
   }
   \put(10,-30){
      \begin{picture}(40,55)(-20,-30)
      \put(-10,-25){\line(0,1){50}}
      \put( 10,-25){\line(0,1){50}}
      \triad{$\bar\tau$}{$\bar z$}
      \put(-10,-27){\makebox(0,0){$\bar z=0$}}
      \put( 10,-27){\makebox(0,0){$\bar z=\beta$}}
      \end{picture}
   }
   \put(-25,-33){\makebox(0,0){(a)}}
   \put( 30,-33){\makebox(0,0){(b)}}
   \end{picture}
}

\def\FigEight{
   \begin{picture}(100,60)(-55,-35)
   \put(-50,-30){
      \triad{$\bar\tau$}{$\bar z$}
   }
   \put(-43,-25){
      \begin{picture}(12,50)(-6,-25)
      \put(-6,-25){\line(0,1){50}}
      \put( 6,-25){\line(0,1){50}}
      \put( 0,-15){\circle*{2}}
      \put( 0, 15){\circle*{2}}
      \put( 0,-18){\makebox(0,0){$\OpB$}}
      \put( 0, 18){\makebox(0,0){$\OpA$}}
      \end{picture}
   }
   \put(-16,-25){
      \begin{picture}(12,50)(-6,-25)
      \put(-6,-25){\line(0,1){50}}
      \put( 6,-25){\line(0,1){50}}
      \put( 0,-15){\circle*{2}}
      \put( 0, 15){\circle*{2}}
      \put( 0,-18){\makebox(0,0){$\OpB'$}}
      \put( 0, 18){\makebox(0,0){$\OpA'$}}
      \put( 0,-15){\vector(0,1){16}}
      \put( 0,  1){\line(0,1){14}}
      \end{picture}
   }
   \put( 11,-25){
      \begin{picture}(12,50)(-6,-25)
      \put(-6,-25){\line(0,1){50}}
      \put( 6,-25){\line(0,1){50}}
      \put( 0,-24){\line(0,1){48}}
      \put( 0,  0){\vector(0,1){0}}
      \put(-6,-25){\line(1,0){12}}
      \put(-6, 25){\line(1,0){12}}
      \end{picture}
   }
   \put( 38,-25){
      \begin{picture}(12,50)(-6,-25)
      \put(-6,-25){\line(0,1){50}}
      \put( 6,-25){\line(0,1){50}}
      \put(-4,-22){\line( 0, 1){36}} \put(-4, -4){\vector( 0, 1){0}}
      \put(-4,-22){\line( 1, 1){8}}  \put( 0,-18){\vector(-1,-1){0}}
      \put( 4, 22){\line( 0,-1){36}} \put( 4,  4){\vector( 0,-1){0}}
      \put( 4, 22){\line(-1,-1){8}}  \put( 0, 18){\vector( 1, 1){0}}
      \end{picture}
   }
   \put(-36,-33){\makebox(0,0){(a)}}
   \put( -9,-33){\makebox(0,0){(b)}}
   \put( 18,-33){\makebox(0,0){(c)}}
   \put( 45,-33){\makebox(0,0){(d)}}
   \end{picture}
}

\def\mdebye{m_{\rm d}}
\def\nf{n_{\rm f}}
\def\LE{{\cal L}_{\scriptscriptstyle\rm E}}
\def\Leff{{\cal L}_{\rm eff}}
\def\im{{\rm Im}\,}
\def\re{{\rm Re}\,}
\def\tr{{\rm tr}}
\def\ca{C_{\rm A}}
\def\tf{t_{\rm F}}
\def\eps{\epsilon}
\def\EulerGamma{\gamma_{\scriptscriptstyle E}}
\def\Ao{A_0}			

\def\OpA{F}
\def\OpB{G}
\def\TC{{\cal TC}}
\def\Rz{{\cal R}_{\bar z}}
\def\polyakov{L}		
\def\mix{\varepsilon}

\def\qsim{ \,\, \vcenter{\hbox{$\buildrel{\scriptstyle ?}\over\sim$}} \,\,}

\let\v=\undertilde

\begin {document}

\preprint {UW/PT-95-06}

\title  {The non-Abelian Debye screening length beyond leading order}

\author {Peter Arnold and Laurence G.~Yaffe}

\address
    {%
    Department of Physics,
    University of Washington,
    Seattle, Washington 98195-1560
    }%
\date {\today}

\maketitle
\vskip -20pt

\begin {abstract}%
    {%
    {%
    \advance\leftskip  -2pt
    \advance\rightskip -2pt
    In quantum electrodynamics, static electric fields are screened at
    non-zero temperatures by charges in the plasma.
    The inverse screening length, or Debye mass,
    may be analyzed in perturbation theory and is of order $eT$
    at relativistic temperatures.
    An analogous situation occurs when non-Abelian gauge theories
    are studied perturbatively, but the perturbative analysis breaks
    down when corrections of order $e^2 T$ are considered.
    At this order, the Debye mass depends on the non-perturbative
    physics of confinement, and a perturbative ``definition'' of
    the Debye mass as the pole of a gluon propagator
    does not even make sense.
    In this work, we show how
    the Debye mass can be defined
    non-perturbatively in a manifestly gauge invariant manner
    (in vector-like gauge theories with zero chemical potential).
    In addition, we show how the $O(e^2 T)$ correction could be determined
    by a fairly simple, three-dimensional, numerical lattice
    calculation of the perimeter-law behavior of large, adjoint-charge
    Wilson loops.
    }%
    \ifpreprintsty
    \thispagestyle {empty}
    \newpage
    \thispagestyle {empty}
    \vbox to \vsize
	{%
	\vfill \baselineskip .28cm \par \font\tinyrm=cmr7 \tinyrm \noindent
	\narrower
	This report was prepared as an account of work sponsored by the
	United States Government.
	Neither the United States nor the United States Department of Energy,
	nor any of their employees, nor any of their contractors,
	subcontractors, or their employees, makes any warranty,
	express or implied, or assumes any legal liability or
	responsibility for the product or process disclosed,
	or represents that its use would not infringe privately-owned rights.
	By acceptance of this article, the publisher and/or recipient
	acknowledges the U.S.~Government's right to retain a non-exclusive,
	royalty-free license in and to any copyright covering this paper.%
	}%
    \fi
    }%
\end {abstract}
\thispagestyle {empty}

\newpage


\section {Introduction}

\begin {figure}
\vbox
   {%
   \begin {center}
      \leavevmode
      
      \epsfbox [150 300 500 500] {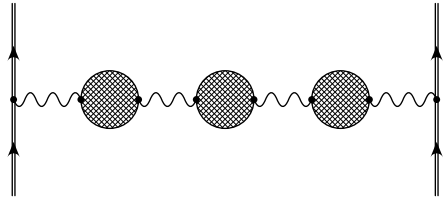}
   \end {center}
   \caption
       {%
       A single virtual photon exchanged between two static test charges.
       \label{figa}
       }%
   }%
\end {figure}

\begin {figure}
\vbox
   {%
   \begin {center}
      \leavevmode
      
      \epsfbox [150 350 500 500] {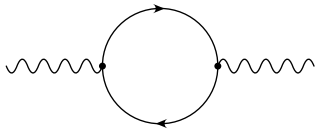}
   \end {center}
   \caption
       {%
       One-loop self-energy of a photon.
       \label{figb}
       }%
   }%
\end {figure}

    Electrically charged particles in a hot plasma
react to electromagnetic fields and cause screening of static electric fields
at large distances.
The inverse screening length, also known as the Debye mass $\mdebye$,
may be computed in QED by considering the exchange of a
single virtual photon between two static test charges,
as depicted in Fig.~\ref{figa}.
\noindent
The long distance fall-off of the static potential
is determined by the position of the
pole in the photon propagator at zero frequency.
This is given by the solution
$\v p^2 = -\mdebye^2$ to
\begin {equation}
   \v p^2 + \Pi_{00}(0,\v p) = 0 \,,
\label{pole def}
\end {equation}
where $\Pi_{\mu\nu}(p_0,\v p)$ is the self-energy of the photon.
In the ultrarelativistic limit (when particle masses and chemical
potentials are negligible),
the leading-order result is easily computed from the
one-loop graph of Fig.~\ref{figb} and yields
\begin {equation}
   \mdebye = {e T \over\sqrt3} + O(e^2 T)
\end {equation}
for a theory with a single fermion of charge $e$.%
\footnote{%
   For reviews, see refs.~\cite{gpy,kapusta}.
}

\begin {figure}
\vbox
   {%
   \begin {center}
      \leavevmode
      
      \epsfbox [150 385 500 440] {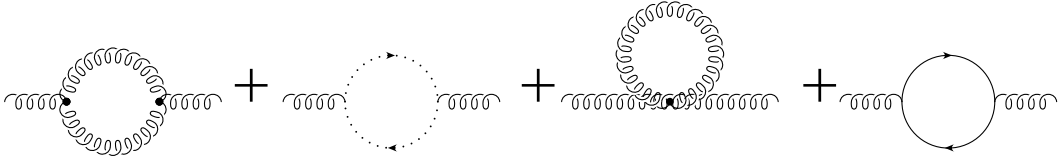}
   \end {center}
   \caption
       {%
       One-loop self-energy of a gluon.
       \label{figc}
       }%
   }%
\end {figure}

Unlike electric fields,
magnetic fields are unscreened, which is reflected by the fact that
\begin {equation}
   \lim_{\v p \to 0} \> \Pi_{ij}(0,\v p) = 0 \,,
   \qquad i,j = 1,2,3.
\end {equation}

A similar perturbative calculation
carried out in non-Abelian gauge theories, using the one-loop graphs
of Fig.~\ref{figc},
yields a lowest-order result of
\begin {mathletters}
\begin {eqnarray}
   \mdebye &=& m_0 + O(g^2 T) \,,
\\
\noalign {\hbox {where}}
   m_0 &=& \left({N\over 3} + {\nf\over6} \right)^{1/2} g T
\end {eqnarray}
\label{m0 eqn}%
\end {mathletters}%
for SU($N$) gauge theory with $\nf$ Dirac fermions.

For the sake of better understanding the nature and reliability of
perturbation theory at finite temperature, there has long been an
interest in computing the leading correction to this
result
\cite{early}.
It is known, however, that this correction
cannot be computed perturbatively in non-Abelian gauge theories
\cite{nadkarni,rebhan}.
As we shall review below, the $O(g^2 T)$ correction to the Debye mass
receives contributions from fundamentally non-perturbative physics
associated with the interactions, at high temperature, of magnetic gluons
with momenta of order $g^2 T$.
The best that can be done perturbatively
is the extraction of a logarithm at that order
\cite{rebhan,braatenP}:
\begin {equation}
   \mdebye = m_0 + {1\over4\pi} N g^2 T \ln\left( m_0 \over g^2 T \right)
                   + c \, g^2 T + O(g^3 T) \,.
\label {c def}
\end {equation}
The constant $c$, however, is not computable by perturbation theory.

   Because the physics of the $O(g^2 T)$ correction is non-perturbative, it
behooves us to formulate a non-perturbative definition of what we mean by the
Debye mass in the first place.  Such a definition should be gauge invariant
and preferably implementable in numerical lattice simulations.
The definition
(\ref{pole def}) is unfortunate in both these respects.
In particular, the self-energy $\Pi_{\mu\nu}$ is not itself
gauge invariant in non-Abelian theories.
There are formal
proofs that the pole position is gauge invariant order by order in
perturbation theory
\cite{kobes},
but this is of limited use since perturbation theory
breaks down beyond leading order.
We should look instead for a definition
that is {\it manifestly} gauge invariant.

   The purpose of this paper is two-fold: ({\em i}) to give a
natural non-perturbative definition of the
Debye mass, and ({\em ii}) to show how the constant $c$ in the expansion
(\ref{c def}) could be extracted from a
relatively simple numerical computation of the perimeter law fall-off of
large, adjoint-charge Wilson loops in three-dimensional, zero-temperature,
pure lattice gauge theory.
In section \ref{eff theory section}, we briefly review the source of
the breakdown of perturbation theory.
In section \ref{defn section}, we construct a
manifestly gauge
invariant, non-perturbative definition of the Debye mass.
We review why one method sometimes suggested in the
literature---extracting the Debye mass from the long-distance
correlation of Wilson lines---is inadequate.
Our definition works only for vector-coupled gauge theories, such as
QCD or QED, and only at zero chemical potential.  We explain what
the difficulties are for axially-coupled theories or non-zero chemical
potentials, and we outline the problems with making a non-perturbative
definition of the Debye mass in those cases.
Finally, section \ref{NLO section} contains our derivation of the
$O(g^2 T)$ correction to the Debye mass
in terms of three-dimensional Wilson loops.


\section{Non-perturbative High Temperature Physics}
\label{eff theory section}

   The physical picture behind our definition will be clearer if we
first review the source of non-perturbative effects in hot non-Abelian
gauge theories.%
\footnote{%
   More details can be found in the discussions and reviews of
   refs.~\cite{gpy,svetitsky,braatenF,shameless}.
}
The problem is easiest to understand by considering a
series of effective theories corresponding to larger and larger distance
scales in the hot plasma.  Since we are interested in studying the
screening of {\it static} electric fields, we can work
directly in Euclidean space where non-zero temperature
corresponds to making the Euclidean time direction periodic with
period $\beta = 1/T$.  So
\begin {equation}
      Z = \int [{\cal D}\bar\psi\, {\cal D}\psi\, {\cal D}A]
          \exp\left(-{1\over g^2}\int\nolimits_0^\beta d\tau\, d^3x\>
                                     \LE \right) \,,
\end {equation}
where we have suppressed details of ghosts and gauge-fixing.
Boson (fermion) fields have (anti-)periodic boundary conditions in Euclidean
time.  At distances large compared to $\beta$, the dynamics of the time
direction decouples, and one obtains an effective {\it three}-dimensional
theory of the zero-frequency modes of the original fields.  Since the
fermionic fields are anti-periodic, only the bosonic degrees of freedom
are relevant in this effective theory.%
\footnote
    {%
    Physically, light bosons dominate over light fermions at low frequency
    because
    of the infrared divergence of the Bose distribution
    $1/(e^{\beta E}{-}1)$ as $E{\to}0$.
    }
Schematically,
\begin {equation}
    Z \to \int [{\cal D}A] \exp\left(-{1\over g^2 T}\int d^3x\>
                                      \Leff\right) \,.
\label{3d theory}
\end {equation}
The effective theory (\ref{3d theory})
is a three-dimensional gauge field $\v A$
coupled to a three-dimensional adjoint-charge scalar
corresponding to $\Ao$.%
\footnote{%
   Purists may object to saying that the adjoint scalar in the effective
   three-dimensional theory arises from $A_0$ in the four-dimensional
   theory, because this statement is gauge dependent.
   The gauge independent identification is that the
   three-dimensional scalar corresponds to the traceless part of the
   path ordered exponential
   ${\cal P} \exp i \int_0^\beta \Ao(\tau,\v x) \> d\tau$
   in the original four-dimensional theory.
   However, we shall continue to refer to the adjoint scalar simply
   as $\Ao$.
}
One of the
effects of integrating out physics with momenta of order $T$ is that the
adjoint scalar obtains a mass of order $g T$.
As indicated in (\ref{3d theory}),
the gauge coupling constant in the three-dimensional theory is
$g_3^2 \equiv g^2 T$.

Next, consider distances in the three-dimensional theory that are large
compared to $1/gT$.
At these distances the adjoint scalar decouples, and the new effective
theory is a pure gauge theory in three dimensions.%
\footnote{
   There can be scalars too if they are part of the fundamental theory,
   such as for electroweak theory, and if the temperature is fine-tuned
   to be near a phase transition.
}
Three dimensional
non-Abelian gauge theories are confining.
Moreover, the only remaining parameter of the theory is the
three-dimensional coupling $g^2 T$ and so, by dimensional analysis, the
confinement radius is of order $1/g^2 T$.  The physics of magnetic gluons with
momenta of order $g^2 T$ is therefore non-perturbative, and the physical
states of the three-dimensional effective theory are glue-balls rather than
individual gluons.  This is unlike the case in zero-temperature {\it
four}-dimensional theories, where the confinement radius diverges
exponentially as $g \to 0$ and non-perturbative contributions are
never the same order as perturbative ones.

It is important to keep in mind that the physics at large distances is the
physics of three-dimensional {\it confinement}, and that it is this
confinement which cuts off infrared divergences encountered in perturbation
theory.  Some papers in the literature work under the misapprehension
that the infrared physics is instead cut off by some sort of mass of order
$g^2 T$ for the gauge field $\v A$.  This is as misleading as thinking
of confinement in zero-temperature QCD as being described by a gluon mass.
A mass would cause large (fundamental-charge) spatial Wilson loops in
high-temperature gauge theory to have perimeter-law behavior because it
would screen the gauge force.
Instead, such loops will have area-law behavior---the signal of confinement.

\section {Defining the Debye Mass}
\label {defn section}

   Begin by considering QED.  The Debye mass can
simply be defined by the correlation length of the equal-time electric
field correlation function:
\begin {mathletters}
\begin {equation}
   	< \v E(\v x) \cdot \v E(0) > \sim
	e^{-\mdebye |\v x|} / |\v x|^3
	\qquad \hbox {as $|\v x| \to \infty$,}
\end {equation}
or
\begin {equation}
	\mdebye \equiv - \lim_{|\v x| \to \infty} \>
	|\v x|^{-1} \ln < \v E(\v x) \cdot \v E(0)> \,.
\end {equation}%
\label {EE def}%
\end {mathletters}%
This is equivalent to the definition (\ref{pole def}) in terms of the
photon self-energy because the exponential rate of decay of a propagator
at large distance is determined by the location of
singularities nearest the real axis in momentum space.%
\footnote
    {%
    The fact that (\ref {EE def}) specifies coincident times
    while (\ref {pole def}) refers to zero frequency
    makes no difference.
    All contributions
    from the (discrete) non-zero frequencies to the equal time
    correlation function decay as $O(e^{-2\pi T |\u x|})$ or faster;
    hence the zero frequency component dominates at large distance.
    }

Unfortunately, this is a poor definition in non-Abelian theories
because $\v E$ is no longer gauge invariant.
One might instead consider a definition in terms of the
correlation between two static test charges.  Specifically, a manifestly
gauge-invariant possibility would be to define the Debye screening length
as the correlation length between Wilson lines (also known as Polyakov loops),
\begin {equation}
	< \polyakov(\v x) \polyakov^@(0) > \qsim
	e^{-\mdebye |\v x|} / |\v x|
	\qquad \hbox {as $|\v x| \to \infty$} \,,
\label{Wilson falloff}
\end {equation}
where the Wilson line
\begin {equation}
  \polyakov(\v x) \equiv
  \tr \> {\cal P} \exp i \int\nolimits_0^\beta \Ao(\tau,\v x) \> d\tau
\end {equation}
is the trace (in the fundamental representation)
of the path-ordered exponential of the line integral of the gauge
field around the periodic Euclidean space.
(${\cal P}$ denotes path ordering.)

\begin {figure}
\vbox
   {%
   \begin {center}
      \leavevmode
      
      \epsfbox [150 290 500 500] {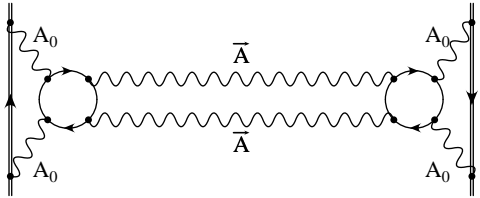}
   \end {center}
   \caption
       {%
        Power-law interaction between two Wilson lines, representing static
	test charges, due to the exchange of a pair of unscreened, magnetic
	photons.
	\label{figd}
       }%
   }%
\end {figure}

\begin {figure}
\vbox
   {%
   \begin {center}
      \leavevmode
      
      \epsfbox [150 350 500 480] {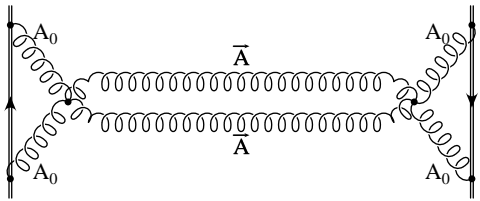}
   \end {center}
   \caption
       {%
	An exchange of magnetic gluons between two Wilson lines.
	\label{fige}
       }%
   }%
\end {figure}

Although this definition has
occasionally been suggested in the literature, it is wrong.
Even in QED, it fails to isolate
the quantity one wants to identify as the Debye mass.
Though Wilson lines
couple directly only to electric fields, they couple {\it indirectly} to
magnetic fields through interactions, and magnetic fields
are not screened.
Fig.~\ref{figd} shows how two Wilson lines can exchange a pair of
magnetic photons in QED, and so, despite the screening of electric fields,
the correlation (\ref{Wilson falloff}) falls off algebraically instead
of exponentially.%
\footnote
    {%
    The potential between arbitrarily heavy test charges,
    $V(\v x) = - \beta^{-1} \ln \> <\polyakov(\v x) \polyakov^@(0) >$,
    decreases as $|\v x|^{-6}$,
    reflecting a magnetic Van der Waals interaction between
    the two electron-positron clouds screening the test charges.
    An analogous case of algebraic screening in non-relativistic theories
    at finite density is discussed in ref.~\cite{cornu}.
    }
In non-Abelian gauge theory, the coupling to the spatial gauge field
can be even more direct
\cite{nadkarni,braatenP},
as in Fig.~\ref{fige}.  The
non-Abelian case is slightly different from QED, however, because
three-dimensional confinement implies that the
Wilson lines cannot exchange a {\it massless}
pair of magnetic gluons; the pair will
instead form a glueball with a mass of order $g^2 T$, and so the
correlation length of Wilson lines defined by
(\ref{Wilson falloff}) will be of order $1/g^2 T$
\cite{nadkarni,braatenP}.
Regardless, this is not the physics of electric screening.

Fortunately, there is a simple symmetry which can be used to exclude the
unwanted exchange of a pair of magnetic photons or a magnetic glueball:
Euclidean time reflection.
Euclidean time reflection corresponds to what in real-time is called
$\TC$, or time reversal times charge conjugation,%
\footnote{
   A mnemonic for this fact: In real time, $\cal CPT$ must be a symmetry
   of any Lorentz invariant (and unitary) theory.  In Euclidean space,
   ${\cal P} {\cal R}_\tau$, where ${\cal R}_\tau$ denotes
   Euclidean time reflection, is a pure rotation and
   must be a symmetry of any 
   Euclidean invariant theory.
   So ${\cal R}_\tau$ must correspond to $\TC$,
   since $\cal P$
   is time independent.
}
and the crucial property
is that $\Ao$ is intrinsically odd under this symmetry while the
spatial gauge field $\v A$ is even.
The Euclidean description is
more convenient for our purposes, and we shall frequently refer to the
symmetry simply as ``time reflection.''
(The reader should note that
in Euclidean functional integrals
time reflection is no more subtle a symmetry than spatial reflection;
there is no extra complication associated with anti-unitarity.)
Euclidean time reflection is a useful symmetry because,
in the effective three-dimensional theory, the {\it only} effect it has
is to negate the adjoint scalar $\Ao$.

If one considers the correlation of a pair of time-reflection odd
operators, instead of the Wilson lines, then the zero-frequency magnetic
contributions of the type depicted in figs.~\ref {figd} and \ref {fige}
will be eliminated.
The lightest intermediate states which can contribute will
be those containing a single $\Ao$ (plus surrounding glue),
so that the correlation length will be $m_0 + O(g^2 T)$.
Any local, gauge-invariant, time-reflection odd operator can be considered
as a replacement for the Wilson line, leading to a general definition of
the Debye mass:

\begin {quote}
  {\it Definition.}  Consider the correlation lengths defined by the
  fall-off, at large spatial separation, of 
  the correlation $<A(\v x) B(0)>$ between operators $A$ and $B$
  that are local (in 3-space), gauge invariant, and odd under
  Euclidean time reflection ({\it i.e.}, real-time $\TC$).
  The inverse Debye screening mass $1/\mdebye$ is the
  largest such correlation length.
\end {quote}

   We are thus able to define the Debye mass directly in terms of the
long-distance fall-off of certain correlation functions.  This definition
will only work, however, in theories where real-time $\TC$ is a good
symmetry; otherwise, there is nothing to prevent states with a single
$\Ao$ from mixing with $\v A$ glueballs, and all of our inverse correlation
lengths will again be $O(g^2 T)$ instead of $O(g T)$ and will
be unrelated to the physics of electric screening.  The restriction to
$\TC$-conserving theories means that the Debye mass cannot be rigorously
defined by the long-distance fall-off of correlation functions in
theories with axial couplings, such as electroweak theory, or in the
presence of a non-zero chemical potential.
We shall comment again on these
cases later, but for now our discussion will be restricted to
vector-coupled theories, such as QCD, at zero chemical potential.

\makeatletter		
\@floatstrue
\def\figure{\let\@capwidth\columnwidth\@float{figure}}
\let\endfigure\end@float
\@namedef{figure*}{\let\@capwidth\textwidth\@dblfloat{figure}}
\@namedef{endfigure*}{\end@dblfloat}
\makeatother

\begin {figure}
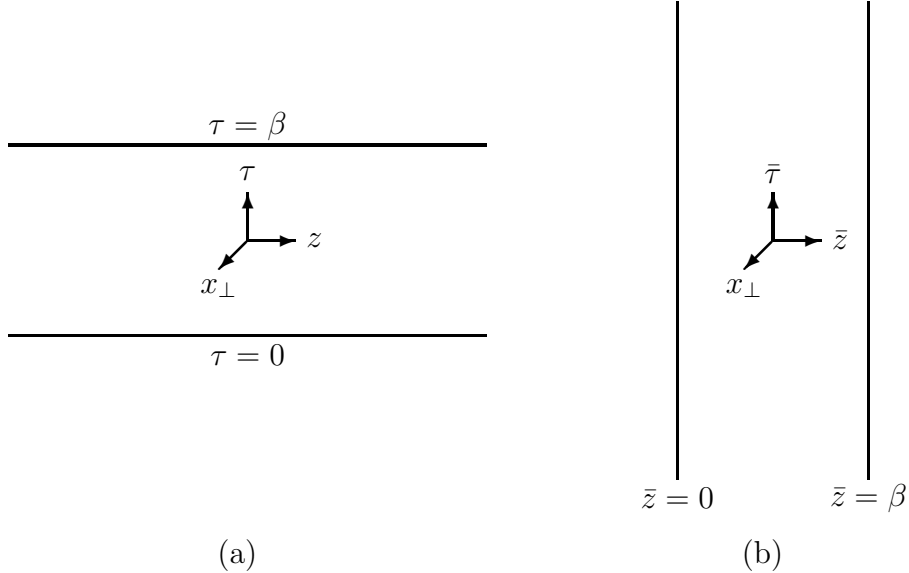

    \FigSix
    \vskip 0.2in
    \caption{
	Interpreting Euclidean time as a periodic spatial direction,
	by relabeling coordinates $(\tau,z)$ as $(\bar z, \bar\tau)$.
	\label{figf}
    }
\end{figure}

Before proceeding further,
it will be convenient to rephrase our definition in alternative language.
Suppose that the separation $\v x$ of our operators is in the $z$ direction.
In Euclidean space, there is nothing that distinguishes the time dimension
as fundamentally different from the spatial ones.
One may turn one's head on the side and interchange
the labels $z$ and $t$, as depicted in Fig.~\ref{figf}.
One then interprets the original four-dimensional
field theory as a zero-temperature theory with one periodic
spatial dimension, instead of a finite-temperature field theory
with all spatial dimensions infinite.
Our correlation functions are now correlation functions
with large separations in ``time,''
and their exponential fall-off is determined by
the energies of the physical states in this zero-temperature,
spatially-periodic field theory.
So the following is an exactly equivalent definition:

\begin {quote}
   {\it Alternative definition.}  Recast the theory as a 3+1 dimensional
   field theory at zero temperature, where one of the spatial
   dimensions---call it $\bar z$---is periodic with period $\beta$.
   Then, in a Hilbert space interpretation,
   the Debye mass is the energy of the lightest
   state that is odd under $\bar z$ reflection.
\end {quote}

For future reference, we should clarify that we will always use the notation
$\Ao$ to denote the component of the gauge field in the periodic direction,
regardless of whether we are interpreting Euclidean ``time'' according
to the original definition or the alternative.


    In addition to their behavior under $\bar z$ reflection,
henceforth denoted $\Rz$,
eigenstates of the spatially-periodic 3${+}$1 dimensional theory
may also be classified according to their behavior under
other space-time symmetries.
Specifically, excitations at rest ($p_x = p_y = 0$)
can be assigned quantum numbers $J^{C}$, $|p_{\bar z}|$, and $R_{\bar z}$
where $J$ is the angular momentum in the $xy$-plane,
$C$ is charge conjugation,
$p_{\bar z}$ is the momentum in the periodic direction $\bar z$, and
$R_{\bar z}$ is the sign acquired under $\bar z$ reflection. 
For $J=0$, there is also one additional quantum number: the sign $P$ of the
state under two-dimensional reflections.%
\footnote{%
   The representations of O(2) = ${\rm Z}(2)\times {\rm SO}(2)$ are
   ({\em i}) two-dimensional representations for each non-zero value of $J^2$,
   and ({\em ii}) two one-dimensional representations,
   distinguished by their Z(2) charge, for $J^2=0$.
}
The lowest energy states will have $p_{\bar z}=0$, and our definition of the
Debye mass restricts us to $R_{\bar z}=-$;
so the states of interest can be summarized by $J^{C(P)}$.
It is not clear {\em a priori} which $J^{C(P)}$ sector
will contain the lightest $R_{\bar z}$ odd state.

\begin {figure}
\begin {table}
\begin {center}
\tabcolsep 10pt

\begin {tabular}{|l|ll|}
\hline
   $J^{C(P)}\kern -5pt $&\multicolumn{2}{l|}
   {\qquad\qquad $R_{\bar z}$ odd operators}
\\
\hline
   $0^{-+}$
   &$\tr \left( \im \! \Omega \right) \equiv \im \polyakov$
   &or \quad
   $\tr \left( F_{03} \> ( F_{12} )^2 \right)$
\\ $0^{+-}$
   &$\tr \left( \im \! \Omega \, F_{12} \right)$
   &or \quad
   $\tr \left( F_{03} \, F_{12} \right)$
\\ $0^{--}$
   &$\tr \left( \im \! \Omega \, [ F_{12} , \, ( F_{i3} )^2 ] \right)$
   &or \quad
   $\tr \left( F_{03} \> [ F_{12} , \> ( F_{i3} )^2 ] \right)$
\\ $0^{++}$
   &$\tr \left( \im \! \Omega \, [ F_{12}, \, [ F_{13}, F_{23} ]]\right)$
   &or \quad
   $\tr \left( F_{03} \> [ F_{12}, \, [ F_{13}, F_{23}]] \right)$
\\ $1^{+}$
   &$\tr \left( \im \! \Omega \> F_{i3} \right)$
   &or \quad
   $\tr \left( F_{03} \, F_{i3} \right)$
\\ $1^{-}$
   &$\tr \left( \im \! \Omega \, \{ F_{12}, F_{i3} \} \right)$
   &or \quad
   $\tr \left( F_{03} \> \{ F_{12}, F_{i3} \} \right)$
\\
\hline
\end {tabular}
\end {center}
\caption
    {%
    Examples of gauge-invariant
    Euclidean time-reflection (or $\Rz$) odd operators
    which couple to specific $J^{C(P)}$ sectors.
    Here,
    $i = 1,2$ is a $2{+}1$ dimensional spatial index and
    $\Omega = {\cal P} \, \exp \int_0^\beta A_0 \, dx^0$
    denotes the un-traced Wilson line or Polyakov loop.
    $\im \! \Omega$ is shorthand for the anti-Hermitian
    part, $\im \! \Omega \equiv (\Omega - \Omega^@)/2$.
    }
\label {tab:ops}
\end {table}
\end {figure}

    Gauge invariant operators which couple to specific
$R_{\bar z}$ odd symmetry channels may be easily constructed.
For example, under $\Rz$ reflection
$A_0 \to -A_0$ and $\polyakov \to \polyakov^@$.
Hence, the time-reflection odd part of the Wilson line
is just the imaginary part.
$\im \polyakov$ is also odd under charge conjugation,
but is even under $x$- or $y$-reflections, so the imaginary part
of the Wilson line probes the $0^{-+}$ sector.
Table \ref {tab:ops} illustrates some of the possible gauge
invariant operators which can be used to probe various symmetry channels.
In the language of the effective three-dimensional theory
(now to be regarded as 2+1 dimensional),
each of these operators creates an $\Ao$ accompanied,
because of confinement, by a neutralizing cloud of glue.
The lowest mass state in
each of these channels will have a mass of $m_0 + O(g^2 T)$.
In a direct lattice determination of the Debye mass,
one should in principle check these, and perhaps other, channels
in order to find the lightest state.
(Alternatively, one could consider correlations of
operators with less symmetry.)

In pure SU(2) gauge theory,%
\footnote
    {%
    These comments also apply to the $p_{\bar z}{=}0$ sector of SU(2)
    theories coupled to fermions (but not complex scalars), since
    the fermions are irrelevant.  In addition, they apply to any group
    which, like SU(2), has only real (or pseudo-real) representations.
    }
the operation of charge conjugation is in fact an element
of the gauge group (namely $i \sigma_2$).
Hence, in this theory, any gauge-invariant state must have $C$ even,
and so the possible sectors are restricted to the $J^{+(P)}$ channels.
(Note that the charge conjugation odd operators shown in table \ref {tab:ops}
vanish identically for SU(2), as they must.)

\subsection* {Wilson lines in pure gauge theories}

    In pure gauge SU($N$) theories
(that is, gauge theories without matter fields%
\footnote
    {%
    Or more generally, theories whose gauge group has a non-trivial center
    but all of whose fields transform trivially under the center.
    }%
)
there is one additional subtlety which occurs
with the Wilson line $\im\polyakov$,
or with more complicated operators containing a Wilson line
wrapping around the periodic $\tau$ direction.
A Euclidean pure gauge theory at non-zero temperature
is invariant not only under periodic gauge transformations;
it is also invariant under
non-periodic gauge transformations that globally multiply
the fundamental representation Wilson line $\polyakov$
by an element of the center of the gauge group.%
\footnote{%
   For a review of this symmetry and its role
   at high temperature, see ref.~\cite{svetitsky}.
}
For SU($N$), the center is Z($N$), the $N$-th roots of unity.
This Z($N$) symmetry is spontaneously broken at high temperature,
and there are $N$ different (pure phase) equilibrium states
distinguished by the phase of the Wilson line,
$\arg <\polyakov> = 2\pi k/N$, $k = 0, \ldots, N{-}1$.
Only one of the equilibrium states (the one in which $<\polyakov> \approx 1$)
is invariant under the naive definition of time reflection.
(Each of the other $N{-}1$ equilibrium states is invariant under
a re-defined time reflection which combines the original reflection
with a non-trivial Z($N$) gauge transformation.)

    In order for
    $\im \polyakov$ to probe the Debye mass,
one must work in the single pure phase equilibrium state
which is invariant under (the chosen definition of) time reflection.
Otherwise, time reflection will fail to select the charge-screening
excitations of interest.
However, a gauge theory functional integral which is invariant under
the Z($N$) center symmetry necessarily averages over all $N$
spontaneously broken phases.
Because of this, the
$\im \polyakov$ correlation length, computed with a Z($N$)
invariant functional integral will be $O(g^2 T)$, and have
nothing to do with the real Debye mass.
This can be seen directly from the fact that the
$<\polyakov \polyakov>$ and
$<\polyakov^@ \polyakov^@>$ correlations
vanish by Z($N$) symmetry, and so
$\im \polyakov$ and $\re \polyakov$ will have identical
$O(g^2 T)$ correlation lengths.

    This difficulty does not reflect any inconsistency in our
definition of the Debye mass, because $\im \polyakov$
is not actually gauge invariant under the full gauge group
of pure gauge theories
(which includes Z($N$) center transformations).
Hence, in pure gauge theories, it does not
meet the requirements stated in the definition.

Nevertheless,
one may avoid this difficulty in pure gauge theories,
and obtain an $O(gT)$ correlation length
for $\im \polyakov$, in either of two ways:
change the operator, or change the theory.
Fixing the operator is easy: simply
replace the fundamental representation trace in the definition
of the Wilson line by the trace in some other complex representation
F which is invariant under the center of the group.
For example, in an SU($N$) theory the symmetric tensor product
of $N$ fundamental representations
is suitable
({\em i.e.}, the 10 of SU(3)).
Then
\begin {equation}
  \im \polyakov_{\rm F} \equiv
  \im \tr \> {\cal P}
  \exp i \int\nolimits_0^\beta d\tau \> \Ao^a (\tau,\v x) \> T^a_{\rm F}
\end {equation}
is invariant under Z($N$) transformations and is completely
unaffected by the spontaneous breaking of the Z($N$) center symmetry.
Or, one may use a local operator (of the same symmetry)
which does not involve a Wilson line at all,
such as $\tr \left( F_{03} \, ( F_{12} )^2 \right)$.

Alternatively, one may restrict expectation values
to include only the single equilibrium state with $<\polyakov> {\approx} 1$
by adding an infinitesimal source to the Lagrangian that biases the
system toward the desired Z($N$) sector:%
\footnote
    {%
    The Z($N$) center symmetry is explicitly broken in theories with
    fundamental representation matter fields; hence in such theories
    no explicit symmetry breaking perturbation need be added.
    }
\begin {equation}
      Z_\epsilon =
          \int [{\cal D}\bar\psi\, {\cal D}\psi\, {\cal D}A]
          \exp\left(-{1\over g^2}\int\nolimits_0^\beta d\tau\, d^3x\, \LE
            \> + \> \epsilon \int d^3x \polyakov
          \right) \,,
\end {equation}
and then send $\epsilon$ to zero after the infinite volume limit.
Therefore, the $\im\polyakov$ correlation length $\xi$ may given by
\begin {equation}
      \xi^{-1} =
      -
      \lim_{|\v x| \to 0} \>
      \lim_{\epsilon\to0^+} \>
      \lim_{{\cal V} \to \infty} \;
      |\v x|^{-1} \>
      \ln \left< \im \polyakov(\v x) \im\polyakov(0) \right>_{\epsilon,\cal V}.
\end {equation}
Adding the source term explicitly breaks the Z($N$) center symmetry,
thereby reducing the full gauge symmetry of the theory so that
$\polyakov$ is now fully gauge invariant.
Of more practical use for numerical simulations, one can simply run
simulations in a large enough volume that one finds no jumping between
the different Z($N$) equilibrium states, as measured by $<\polyakov>$.
One would need to
take data generated by a run in a single pure phase
where $\bar\polyakov{\equiv}<\tr\polyakov>$ is non-zero,
and then measure the long-distance fall-off of
\begin {equation}
   <\im[\polyakov(\v x)\bar\polyakov^@] \, \im[\polyakov(0)\bar\polyakov^@]>
\end {equation}
in lieu of $<\im\polyakov(\v x)\, \im\polyakov(0)>$.

\subsection* {Axial theories and chemical potentials}

\begin {figure}
\vbox
   {%
   \begin {center}
      \leavevmode
      
      \epsfbox [150 170 500 420] {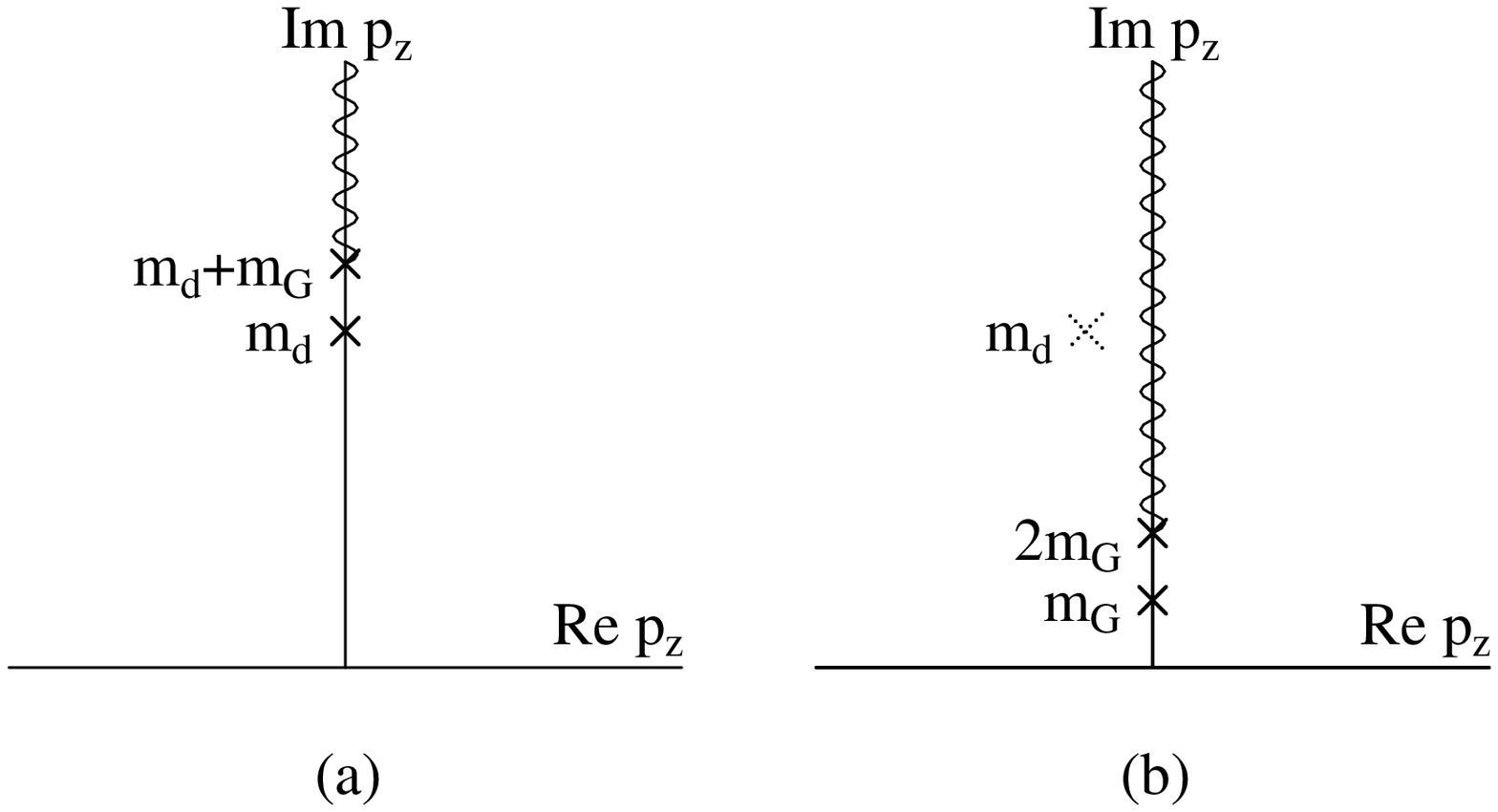}
   \end {center}
   \caption
       {%
       The singularity structure, in the complex $p_z$ plane, of a
       correlation of Euclidean time-reflection odd operators in a
       theory where real-time $\TC$
       ($a$) is, or ($b$) is not, a good symmetry.
       $m_{\rm G}$ is $O(g^2 T)$ and stands for the lightest glueball
       mass in the theory when the $z$ direction is regarded as time.
       For simplicity, we have suppressed all singularities associated
       with excited glueball or excited Debye states.
       \label{fig:singularity}
       }%
   }%
\end {figure}

As mentioned earlier, our definition of the Debye mass
does not work for theories in which
real-time $\TC$ is not a symmetry.
Hence, it cannot be applied to gauge theories with axial couplings,
or in the presence of a non-zero chemical potential.
In the language of our alternative definition, where the
$z$ direction is viewed as ``time,'' the problem manifests as follows:
the lightest state with a single $A_0$ is no longer stable against decay
into an $\v A$ glueball.
Nevertheless,
there is still a singularity in the complex
momentum plane associated with this $A_0$ ``resonance.''  The situation is
depicted in Fig.~\ref{fig:singularity}, where we have considered a generic
correlation of Euclidean time-reflection odd operators and sketched some
features of its singularity structure in the complex $p_z$ plane.
We have assumed $p_\perp = 0$ for simplicity.
The case for a $\TC$
conserving theory is shown in Fig.~\ref{fig:singularity}($a$), where the
location of the singularity closest to the real axis is our Debye mass.
Introducing a small amount of $\TC$ violation will
mix in $\v A$ glueball states, changing
the analytic structure to that of Fig.~\ref{fig:singularity}($b$).
The Debye singularity from Fig.~\ref{fig:singularity}($a$) is still present
but has moved slightly off the axis onto the second sheet.

One can still imagine, in principle,
defining a (complex) Debye mass based on the location of the pole.
This is, in fact, what one does every day when talking about the
mass of an unstable particle such as the $Z$-boson or the $\pi^0$.
There is an important difference, however, which is that the $z$ direction
is not really a time direction and can be considered one only by analytic
continuation.  In the real world, one can reach
the $Z$ resonance experimentally by making
$-p_0^2 + \v p^2$ close to $-M_{\rm Z}^2$.  In contrast,
one cannot experimentally study the static ($p_0=0$) properties
of the plasma by taking $\v p^2$ close to $-\mdebye^2$.
And because the introduction of finite temperature breaks Lorentz
invariance, studying $p_0 \not= 0$ instead is not equivalent:
the physics of the dynamics of real plasma excitations is not the
same as the physics of electric screening.  The moral is that defining
the Debye mass by the location of the relevant singularity in
Fig.~\ref{fig:singularity}($b$) would be somewhat abstract.

Another possible method for defining an electric screening length is in
terms of carefully chosen moments of particular correlation functions.
For the purpose of illustration, assume we had a correlation function
that behaved like
\begin {equation}
   G(r) = {1\over r} \, e^{-a g T r} + {\mix\over r} \, e^{-b g^2 T r} \,.
\label{G example}
\end {equation}
The first term is the behavior we would have in a $\TC$ invariant theory;
the second term represents the mixing with glueball states due to
interactions breaking $\TC$, with $\mix$ the amplitude of that mixing.
Now consider defining a correlation length by the ratio of moments
\begin {equation}
   \xi_G \equiv {\int\nolimits_0^\infty r^2 \, G(r) \, dr
                 \over
                 \int\nolimits_0^\infty r \, G(r) \, dr} \,.
\end {equation}
This will yield
\begin {equation}
   \xi_G = (a g T)^{-1} \times [1 + O(\mix/g^2)] \,.
\end {equation}
As long as $\mix$ is small compared to $g^2$,
this will give a correlation length
of order $1/gT$ that, at leading order, matches what we want
to call the inverse Debye mass.  So one could simply define
what one means by the Debye mass to be precisely $1/\xi_G$.
The problem with a definition of this form is that it is completely
convention dependent; the resulting value depends on exactly which
correlation function and which moments are used in the definition.
In contrast, lengths characterizing the exponential long-distance
decay of correlations, as in our original definition for $\TC$
invariant theories, do not generically depend on the details of the
operators used, other than their symmetries.

If the Debye mass is so difficult to define at finite chemical potential,
what about the case of non-relativistic QED\@?
Why do physicists generally
have few qualms about discussing exponential screening in such plasmas?
The reason is a matter of scale and what one means by ``long-distance.''
Ref.~\cite{cornu} gives a calculation of the charge-charge correlation
in such plasmas and finds that it does fall algebraically at very large
distance.
However, as one increases the distance in a variety of physical
applications, the correlation first
falls exponentially for many $e$-foldings before finally tapering off
in algebraic behavior, and so the concept of exponential screening is
useful in practice.

Now consider the case of relativistic gauge
theories and our toy example (\ref{G example}) of a correlation $G(r)$.
The number of $e$-foldings over which the first term dominates is
only order $\ln(1/\mix)$.
Unlike our toy correlation $G(r)$, a real correlation will have
additional contributions from
{\it excited\/} time-reflection odd states, with energies of
$\mdebye{+}O(g^2T)$:
\begin {equation}
   G(r) \to {1\over r} \sum O(1) \, e^{-[a g T + O(g^2 T)]r}
          + {1\over r} \sum O(\mix) \, e^{-O(g^2 T) r}
          + \hbox{(other junk)}\,.
\end {equation}
In the range where the glueball contributions are small, $r$
is still too small to suppress these excited states
unless $g\ln(1/\mix) \gg 1$.
Therefore, an approximate definition of the Debye mass, in terms of
the intermediate-range fall-off of correlation functions, is not
useful beyond leading order unless the amount of $\TC$ violation is
extraordinarily small.  In particular, it is not useful if the mixing
$\mix$ is simply some power $g^n$ of the coupling.


\section {The {$O(\lowercase{g}^2T)$} correction to the Debye mass}
\label {NLO section}

We return now to vectorially-coupled theories at zero chemical potential.
If one is interested only in the $O(g^2 T)$ correction to the Debye mass,
then it is possible to reduce the computation of the Debye mass to a much
simpler problem than the extraction of correlation lengths in
a four-dimensional theory with a small periodic dimension and
dynamical fermions.
This simplification will emerge from the successive reduction to equivalent
effective theories describing longer distance scales, as discussed in
section~\ref{eff theory section}.
The philosophy is similar to that applied by Braaten \cite{braatenF}
to the expansion of the free energy in powers of $g$.
(With more work, it could be extended to handle even higher-order
corrections to the Debye mass.)
The result, to be derived momentarily, expresses the $O(g^2 T)$ part
of the Debye mass in terms of the perimeter law coefficient of
adjoint-representation Wilson loops in a three-dimensional pure gauge theory.
This relation is particularly nice in that it holds regardless of
which symmetry channel of the 2$+$1 dimensional theory has the lowest mass
time-reflection (or $\Rz$) odd excitations.

\begin{figure}
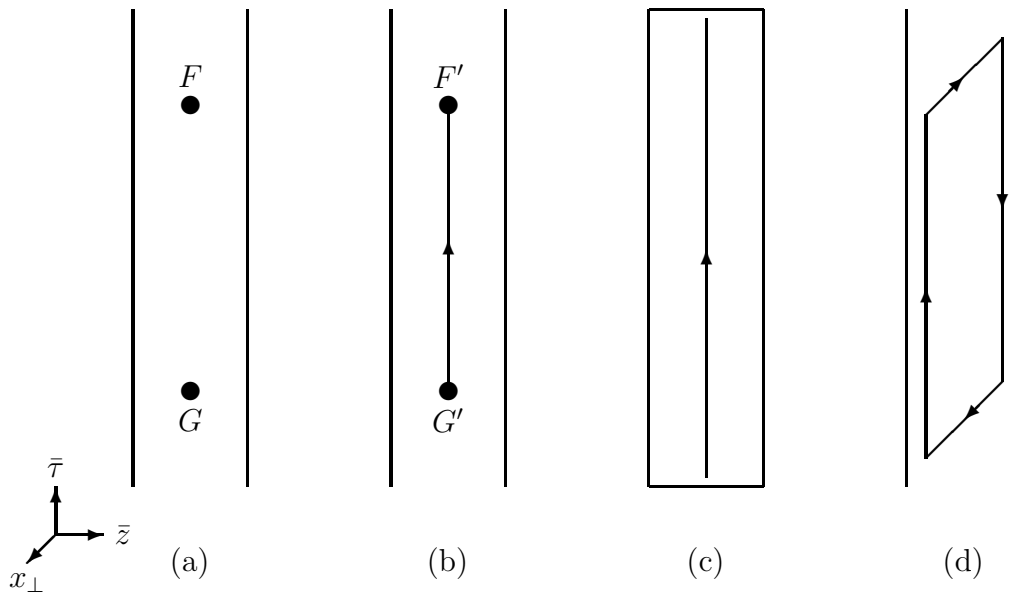

  \vskip -0.0in
  \FigEight
  \vskip 0.2in
  \caption{
   Various stages in the reduction relating the
   $O(g^2 T)$ correction to the Debye mass
   to the perimeter law coefficient of
   three-dimensional adjoint-representation Wilson loops.
   ($a$) The correlation of some pair
   $\OpA$ and $\OpB$ of $\Rz$ odd operators at large separation.
   A three-dimensional reduction is performed and
   only the $p_{\bar z}=0$ modes are relevant to what follows.
   ($b$) The $\Ao$ field is integrated out, generating an $\Ao$
   propagator connecting modified insertions
   $\OpA' \sim \partial \OpA/\partial \Ao$ and
   $\OpB' \sim \partial \OpB/\partial \Ao$.
   Since $\Ao$ is heavy, its propagator may be replaced by
   a straight, adjoint-representation, Wilson line.
   The long-distance fall-off of the correlation measures
   the energy of glue needed to screen a heavy adjoint charge.
   ($c$) If one makes the $\bar\tau$ direction
   periodic with an arbitrarily large period $\Theta$,
   then the minimal screening energy of an adjoint charge
   is determined by the fall-off of an adjoint Wilson line wrapping
   around the $\bar\tau$ direction.
   (This is not to be confused with the original Polyakov loop $L$,
   which wraps around 
   the small $\bar z$ direction.)
   ($d$)
   The fall-off of the long adjoint line with increasing length
   is the same as the perimeter law fall-off of any large,
   adjoint Wilson loop.
   (The loop
   depicted in the figure is meant to be large in $\protect{\v x_\perp}$ as
   well as in $\bar\tau$.)
   \label{fig:reduction}
  }
\end{figure}

First,
reduce the problem to an effective three-dimensional theory
by integrating out modes with non-zero frequency in the
periodic direction.
If we relabel the $z$ axis as ``time,''
we want to know the energy of an $\Ao$, together with its
cloud of glue, propagating forward in time in 2+1 dimensions.
Next, make a further reduction to an effective theory for
distances large compared to $1/gT$,
so that the bare $\Ao$ can now be considered heavy.
The resulting effective theory is simply a 2$+$1 dimensional pure gauge theory
(plus irrelevant corrections suppressed by powers of $g$).
The {\it non-perturbative} contribution
of the cloud of glue surrounding the $\Ao$ is not sensitive to
whether the $\Ao$ is merely heavy or is infinitely heavy.
The propagation of the bare $\Ao$ can then be replaced by an
adjoint-charge Wilson line, exactly analogous to the way in which
extremely heavy quarks in zero-temperature QCD can be replaced by
fundamental-charge Wilson lines.
The non-perturbative piece of the Debye mass is given by the energy
of the glue required to screen an infinitely heavy adjoint charge,
which can be extracted from a numerical lattice calculation of
the perimeter-law behavior of large Wilson loops.
Schematically,
\begin{equation}
   \mdebye = m_{\rm pert} + \Delta m \,,
\label{dm split up}
\end{equation}
where $m_{\rm pert}$ is a perturbative contribution to the mass
and where $\Delta m$ is extracted from
\begin {equation}
   \left<\tr \, {\cal P} \exp\left(
        i \oint\nolimits_C d\v x \cdot \v A_{\rm adj}
        \right)\right>
   \sim \exp[-\Delta m \> {\rm length}(C)] \,,
   \qquad
   \hbox{for large loops $C$,}
\end {equation}
in the three-dimensional gauge theory.
Note that fermions are completely absent from the calculation of $\Delta m$
because they decoupled in the three-dimensional limit.
Figure \ref {fig:reduction} illustrates the various
stages leading to the relation (\ref {dm split up}),
which will be discussed in more detail below.

At each stage in the reduction,
it is necessary to match carefully the effective theories
onto the original four-dimensional theory,
in order to keep track of the perturbative piece
$m_{\rm pert}$ of (\ref{dm split up}) correctly.
Fortunately, at the order of interest,
the matching is fairly simple and straightforward.

\makeatletter		
\@floatsfalse
\def\figure{%
\let\@capwidth\columnwidth
\ifpreprintsty\iffirstfig
{\newpage\centerline{FIGURES}}\global\firstfigfalse
\fi\fi
\vskip1pc
\def\@captype{figure}%
\interlinepenalty10000 %
\@ifnextchar[{\@chuckoptarg}{}%
}%
\def\endfigure{\goodbreak\vskip1pc}%
\@namedef{figure*}{\figure}%
\@namedef{endfigure*}{\endfigure}%
\makeatother

Before we dive into the details of matching, notice that this
picture of the Debye mass makes the presence of the logarithmic
$O(g^2 T \ln(m_0/g^2T))$ correction found in (\ref{c def}),
as well as its coefficient, trivial to understand.
Consider the self-energy of a static, infinitely heavy charge
in 2+1 dimensions.
If we were discussing QED coupled to such a charge,
then the electric field surrounding the charge would be
${\cal E}=e/2\pi r$, and the energy of that electric field would be
\begin {equation}
   E = {\textstyle{1\over2}} \int d^2x \, {\cal E}^2
     = {e^2\over4\pi} \int\nolimits_0^\infty {dr\over r} \,,
\label{E estimate}
\end {equation}
which is logarithmically divergent in both the infrared and the
ultraviolet.  This same picture holds for a non-Abelian gauge theory
at distances small compared to the confinement scale,
since then gluon self-interactions are small.
The confinement scale, however, provides
an infrared cut-off at $r^{-1} {\sim} g^2T$.
The mass of the charged particle, which is order $m_0$ and
not actually infinite, provides the ultraviolet cut-off.
Finally, since the heavy particle is an adjoint charge, $e^2$ should
be replaced by $\ca \, g_3^2$, where $\ca$ is the quadratic Casimir
number for the adjoint representation, or $N$ for SU($N$).
Eq.~(\ref{E estimate}) then precisely reproduces the
logarithm of (\ref{c def}).

We now turn to fleshing out the details of the split up (\ref{dm split up})
and matching of the sequence of effective theories.
At each stage, an effective theory will describe the same long-distance
physics as its shorter-distance predecessor, provided its parameters
are carefully matched to the parameters of its predecessor.
One can achieve this matching by computing and equating a
set of long-distance quantities in both theories.
The required matching reflects the different treatment
of {\it short}-distance physics in the two theories;
it doesn't depend in detail on the physics at long distances,
which in our case is non-perturbative.
Hence, if one temporarily modifies the theories by introducing
a long-distance cut-off, and one uses the same cut-off for both theories,
then the matching of the infrared cut-off theories
will {\it also} provide the correct matching for the theories
when the infrared cut-off is removed.
The temporary introduction of an infrared cut-off is merely a convenience
which allows one to compute the matching perturbatively.


\subsubsection*{Step 1: Reduction to three dimensions}

This reduction has been extensively treated in the literature
\cite{braatenF,3d reduction A,3d reduction B,3d reduction C,3d reduction D}.
At the order we are interested in, the matching is very simple.
Integrating out the non-static ($p_0\ne0$) components of the fields
generates a mass term for $\Ao$ through diagrams such as Fig.~\ref{figc}.
The effective theory is of the form
\begin {equation}
   S_3 = {1\over \bar g^2 T} \int d^3 x \left[ \,
     {\textstyle {1\over4}} F_{ij} F_{ij}
     + {\textstyle {1\over2}} (D_{\rm adj} \Ao)^2
     + {\textstyle {1\over2}} m_0^2 \Ao^2
     + (\hbox{higer-order}) \,
   \right] \,.
\label{S3}
\end {equation}
The ``higher-order'' term denotes marginal and irrelevant operators in the
effective theory which are suppressed by explicit powers of $g$ and whose
effect can be ignored at the order of interest.
The dimensionless coupling $\bar g$ is related to the original
four-dimensional coupling by
\begin {equation}
   \bar g^2 = g^2(T) + O(g^4) \,,
\end {equation}
where evaluating the four-dimensional coupling $g$ at a renormalization
scale of order $T$ eliminates large logarithms from the higher-order
corrections to this matching condition.
The mass $m_0$ is just, up to $O(g^3 T)$ corrections from two loop diagrams
that don't concern us,
the mass (\ref{m0 eqn}) we introduced in the introduction:
\begin {equation}
   m_0^2 = {1\over3} \left(\ca + \sum_{\rm F} \tf\right) g^2 T^2
             + O(g^4 T^2) \,.
\end {equation}
We have been slightly more general than the SU($N$) case of
(\ref{m0 eqn});
$\ca$ is the quadratic Casimir for the adjoint representation
and $\tf$ is the normalization of each irreducible
fermion representation F:
\begin {equation}
   \ca \delta^{ab} = f^{acd} f^{bcd} \,,
   \qquad
   \tf \, \delta^{ab}
   = \tr \left( {\bf T}_{\rm F}^a {\bf T}_{\rm F}^b \right) \,.
\end {equation}

We have not yet specified an ultraviolet renormalization scheme
for the effective theory (\ref {S3}),
because it is irrelevant in this step at the order shown above.
However we will
need a specific scheme in the next step, so let us pick one.  We will use
dimensional regularization in $d=3-2\eps$ dimensions for our
effective theory, with the minimal
subtraction scheme and a renormalization scale $\mu=T$.


\subsubsection*{Step 2: Replacing $\Ao$ by a Wilson line}

Now we want to integrate out $\Ao$ and move to an effective theory
for momenta small compared to $m_0$.  But our goal is to describe the
propagation of an $\Ao$ itself, which we might probe by the long-distance
behavior of some gauge-invariant correlation in the three-dimensional
theory.
For example, consider the three-dimensional analog of the first
$0^{+-}$ operator listed in Table I,
\begin {equation}
    \left <
        \tr\left( \Ao F_{12} \right)_{(0,\v x)} \,
        \tr\left( \Ao F_{12} \right)_{(0,0)}
    \right> \,.
\label {eq:foo}
\end {equation}
Imagine evaluating this correlation by first doing the path integral over
$\Ao$ and only later doing the path integral over $\v A$.
The integral over $\Ao$ will replace the $\Ao$'s appearing
in the integrand by a propagator of the $\Ao$ field in the
background of $\v A$,%
\footnote{%
   This is true in the approximation that we ignore quartic and higher-order
   interactions among $\Ao$ in the effective Lagrangian (\ref{S3}).
   As mentioned before, such terms are suppressed by explicit powers of
   the coupling and do not contribute to the Debye mass at the order we
   are interested.  To go to higher orders, one would have to include
   these terms and treat them as perturbations.
}
which is the solution to
\begin {equation}
   \left( -D^{\rm adj}_i D^{\rm adj}_i + m_0^2 \right) \,
   \Delta_{m_0,\v A}(\v x)
   = \delta^3(\v x) \,.
\label{green function}
\end {equation}
Now suppose that $\v A$ has only low-momentum components and is
smooth on the scale of $1/m_0$.  The solution to (\ref{green function}) can
then be expressed as an expansion in powers of derivatives of $\v A$
and gives
\begin {equation}
   \Delta_{m_0,\v A}(\v x)
   = \Delta_{m_0}(\v x) \; {\cal P}
     \exp\left(i\int\nolimits_0^{\v x} d\v x \cdot \v A_{\rm adj} \right)
     + O(\partial \v A) \,,
\label{wilson sub}
\end {equation}
where the integration path is a straight line from the origin to
$\v x$, and where $\Delta_{m_0}$ is the free scalar propagator
which behaves as
\begin {equation}
   \Delta_{m_0}(\v x) \sim {e^{-{m_0}|\v x|} \over4\pi |\v x|}
   \qquad
   \hbox{for $|\v x| \to \infty$} \,.
\end {equation}
Hence,
the net effect of integrating out $\Ao$
will be to replace
the pair of $\Ao$'s in a correlation such as (\ref {eq:foo})
by an adjoint-representation path-ordered exponential,
times an overall factor:
\begin {equation}
   <\Ao(\v x) \Ao(0) \cdots>
   \to
   e^{-m_1|\v x|} \left< {\cal P}
     \exp\Bigl(i\int\nolimits_0^{\v x} d\v x \cdot \v A_{\rm adj} \Bigr)
     \cdots \right>.
\label {eq:sub}
\end {equation}
This corresponds to Fig.~\ref {fig:reduction}($b$),
and naively $m_1$ is just $m_0$.
The substitution (\ref {eq:sub})
will be valid for separations large compared to the inverse
Debye mass.

But, as sketched in Fig.~\ref{fig:reduction}($c$),
once one introduces the adjoint representation line,
one may dispense with the details of the operator insertions
in the original correlation function by considering the
$\bar\tau$ direction (the $2{+}1$ dimensional ``time'')
to be periodic with an arbitrarily large
period $\Theta$ and computing the expectation of a straight
adjoint Wilson line wrapping around the $\bar \tau$ direction:
\begin {equation}
   e^{-m_1 \, \Theta} \left< \tr \; {\cal P}
     \exp\Bigl(i\int_0^\Theta d\bar\tau \cdot \v A_{\rm adj} \Bigr)
     \right> \,.
\label {straight wilson loop}
\end {equation}
In the Hilbert space interpretation, this corresponds to a trace over
all states containing a single adjoint-representation external charge.
(And includes a sum over all $J^{C(P)}$ sectors.)
The exponential fall-off of this expression with the period $\Theta$
will be determined by the energy of the lightest such state ---
which is precisely our definition of the Debye mass.

Finally, as indicated in Fig.~\ref{fig:reduction}($d$),
the same coefficient for the fall-off of the
correlation with contour length may instead be obtained by considering
a large, topologically trivial, adjoint loop:
\begin {equation}
   e^{-m_1 \, |C|} \left< \tr \; {\cal P}
     \exp\Bigl(i\oint_C d\v x \cdot \v A_{\rm adj} \Bigr)
     \right> \,.
\label {wilson loop}
\end {equation}
The exponential fall-off of this expression with contour length $|C|$
will yield
the Debye screening length through $O(g^2 T)$.
The perimeter-law decay of the
Wilson loop gives the energy of the glue surrounding the heavy adjoint
charge, while $m_1$ above is the mass of the bare charge.

There is just one complication: the
substitution (\ref{wilson sub}) is only a good approximation in the
presence of gauge fields $\v A$ with small momentum.  Our two effective
theories---the one with $\Ao$ and the one where we've replaced it
by a Wilson loop---differ in how they treat large momentum effects.
As usual, this means we need to carefully adjust parameters in order
to make the two theories describe the same long-distance physics.
In particular,
the correct choice of $m_1$ in (\ref{wilson loop}) is not necessarily
$m_0$; it must be determined by matching.
This is done by perturbatively computing the long-distance fall-off
of the $\Ao$ correlation functions in both theories,
which we will only need to do at one-loop order.
Dimensional regularization will be used as our infrared cut-off.

\begin {figure}
\vbox
   {%
   \begin {center}
      \leavevmode
      
      \epsfbox [150 380 500 470] {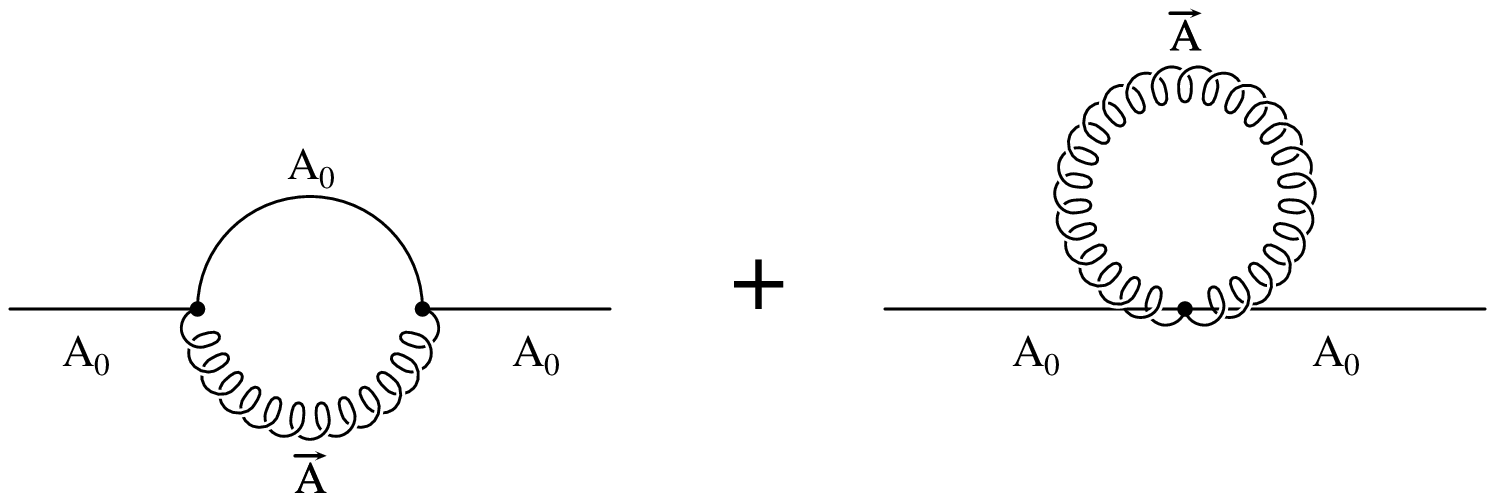}
   \end {center}
   \caption
       {%
	The one-loop self-energy for $\Ao$ in the three-dimensional theory.
	\label{figg}
       }%
   }%
\end {figure}

In the original three-dimensional effective theory,
the fall-off of the $\Ao$ propagator is determined by the
position of the pole.
The shift in the pole position due to the one-loop self energy
can be computed from the diagrams of Fig.~\ref{figg} and gives a
contribution to the Debye mass of
\begin{eqnarray}
   \delta m_0^2
   &=& \ca \, \bar g^2 \mu^{1+2\eps} \int {d^{3-2\eps}q\over(2\pi)^{3-2\eps}}
      \Biggr\{
         {1\over q^2+m_0^2}
         + {(1-2\eps)\over q^2}
         + {2(m_0^2-p^2)\over q^2[(p+q)^2+m_0^2]}
\nonumber
\\
   && \qquad\qquad\qquad\qquad\qquad
         + (\xi-1) (p^2+m_0^2)
            {q^2 + 2p\cdot q \over q^4[(p+q)^2+m_0^2]}
      \Biggl\} \Biggl|_{p^2=-m_0^2} \,,
\\
\noalign {\hbox {which yields}}
   \delta m_0
   &=& {1\over8\pi} \, \ca \, \bar g^2 \mu \left[
       - {1\over\epsilon} + \ln\left(m_0^2 \over \pi\mu^2\right)
       + \EulerGamma - 1 \right] \,.
\label{dm0 eqn}
\end{eqnarray}
We have worked in covariant gauge with gauge parameter $\xi$.

\begin {figure}
\vbox
   {%
   \begin {center}
      \leavevmode
      
      \epsfbox [150 330 500 490] {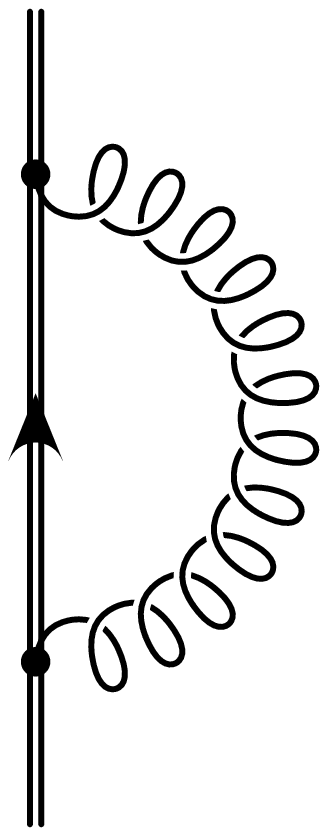}
   \end {center}
   \caption
       {%
        The one-loop self-energy for a Wilson line.
	\label{figh}
       }%
   }%
\end {figure}

Correspondingly, to obtain the fall-off of an adjoint-charge
Wilson line
we need to compute the one-loop self-energy correction
for a static source, as shown in Fig.~\ref{figh}.
Locally, we can treat a very large Wilson loop as straight.
Alternatively, we can confine the system
to a very large but finite volume and let a straight Wilson line wrap
periodically
around the space, as in Fig.~\ref{fig:reduction}($c$).
In any case, taking the line to be in the $z$ direction,
Fig.~\ref{figh} gives
\begin {equation}
   \delta m_1 = \ca \, \bar g^2 \mu \int_0^\infty dz\> \Delta_{zz}(z,\v 0)
   = 0 \,.
\label{dm1 eqn}
\end {equation}
where $\Delta_{ij}(z,\v x_\perp)$ is the $\v A$ propagator and the
integral vanishes in dimensional regularization.

Matching requires that $m_0 + \delta m_0 = m_1 + \delta m_1$.
Putting (\ref{dm0 eqn}) and (\ref{dm1 eqn}) together,
and continuing to choose our renormalization scale in each
effective theory to be $\mu=T$, gives
\begin {equation}
   m_1 = m_0 + {1\over8\pi} \, \ca \, \bar g^2 T \left[
       - {1\over\epsilon} + \ln\left(m_0^2 \over \pi T^2\right)
       + \EulerGamma - 1 \right] \,.
\label {m1 eqn}
\end {equation}
The coupling $\bar g$ here still represents the coupling $g(T)$ in the
original four-dimensional theory.  The fact that the matching of couplings
between effective theories becomes non-trivial beyond leading order
won't enter into our results at the order of interest.


\subsubsection*{Step 3: From the continuum to the lattice}

\begin {figure}
\vbox
   {%
   \begin {center}
      \leavevmode
      
      \epsfbox [150 350 500 480] {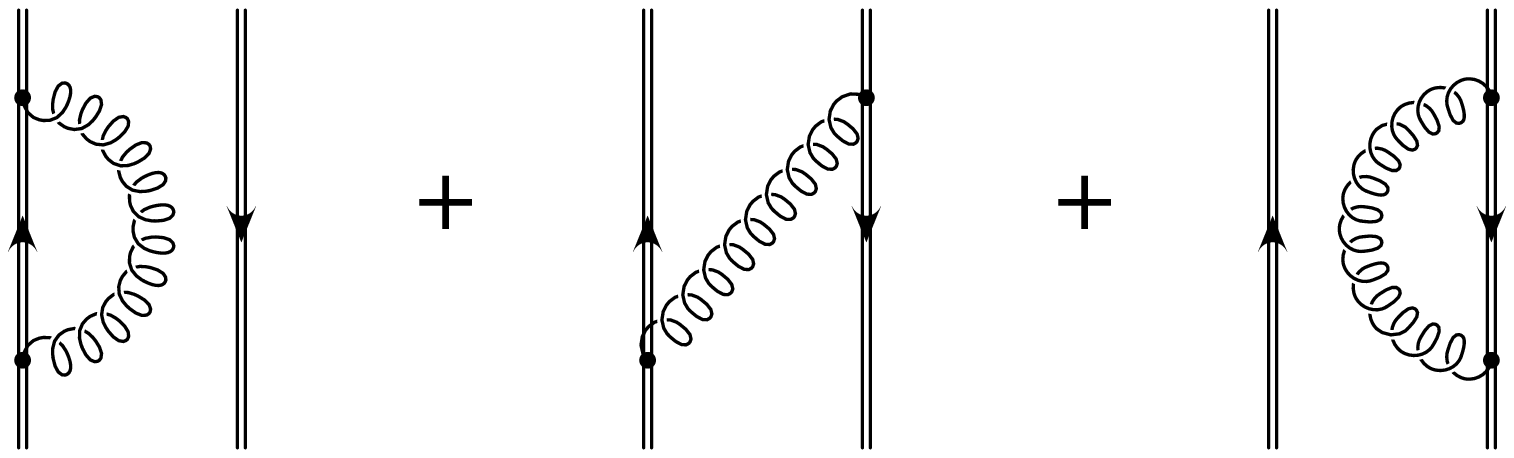}
   \end {center}
   \caption
       {%
        The one-loop self-energy for two opposing Wilson lines, separated
	by a distance $R$.
	\label{figi}
       }%
   }%
\end {figure}

To measure the perimeter-law behavior of large adjoint Wilson loops
numerically, one will put the system on a lattice.%
\footnote{%
   If the lattice theory is defined with link variables $U$ in
   some representation (typically the fundamental representation),
   then the adjoint Wilson loop is given by the
   path-ordered product of $U_{\rm adj}$ over the links of the loop,
   where $U^{ab}_{\rm adj}$ is $\tf^{-1}\, \tr(T^a U T^b U^@)$ for each link.
}
The ultraviolet will be regulated by the lattice spacing instead of by
dimensional regularization, and so we need to modify our
matching condition (\ref{m1 eqn}) appropriately.  We will do this
by again matching the one-loop self-energy, now between an adjoint
line in the continuum and one on the lattice.  However, dimensional
regularization is no longer a good choice for our temporary, common,
infrared cut-off in the two theories.  Instead, we shall consider
two, opposite, parallel Wilson lines running in the $z$ direction and
separated by a large distance $R$ in $x$.  This provides an infrared cut-off
because the lines neutralize each other when viewed from large
$\v x_\perp = (x,y)$.  The one-loop contribution to the energy of
these lines is shown in Fig.~\ref{figi}.

In the continuum, the result is
\begin {eqnarray}
   \delta' m_1
   &=& \ca \, g^2 \mu^{1+2\eps} \int_0^\infty dz \>
     \left[ \Delta_{zz}(z,\v 0) - \Delta_{zz}(z,\v R) \right]
\nonumber\\
   &=& {1\over2} \ca \, g^2 \mu^{1+2\eps}
       \int {d^{2-2\eps}p_\perp \over (2\pi)^{2-2\eps}}
     \> {1\over p_\perp^2} \left(1 - e^{i \v p_\perp \cdot \v R}\right)
\nonumber\\
   &=& {1\over8\pi} \, \ca \, g^2\mu \left[\,
     {1\over\eps} + \ln(\pi\mu^2 R^2) + \EulerGamma \,\right] + O(\eps)
   \,.
\end {eqnarray}
In lattice perturbation theory,
the result is similar but we get lattice propagators
instead of continuum ones:
\begin {eqnarray}
   \delta m_{\rm lat}
   &=& {1\over2} \ca \, g^2 a^{-1} 
     \int {d^2p_\perp \over (2\pi)^2} \>
     \Delta_{\rm lat}(p_\perp) \left(1 - e^{i p_x R/a}\right)
   \,,
\label{lattice integral}
\end {eqnarray}
where $a$ is the lattice spacing; $p_\perp$ is in lattice units and restricted
to the Brillouin zone $|p_x|,|p_y| \lt \pi$; and
\begin {equation}
   \Delta_{\rm lat}(p_\perp)
   = {1\over4} \left[ \, \sin^2{p_x\over2} + \sin^2{p_y\over2} \,\right]^{-1}
   \,.
\end {equation}
The $R \gg a$ limit of the integral is extracted in the appendix and gives
\begin {equation}
   \delta m_{\rm lat} \to {1\over8\pi} \, \ca \, g^2 a^{-1} \left[\,
       \ln\Bigl({8 R^2 / a^2}\Bigr) + 2\EulerGamma \right]
   \,.
\end {equation}
To match the theories, we must pick $a^{-1}=\mu=T$ to make the
coupling constant definition match up, and then require
$m_{\rm lat} + \delta m_{\rm lat} = m_1 + \delta' m_1$, or
\begin {equation}
   m_{\rm lat} =
   m_0 + {1\over8\pi} \, \ca \, g^2 T \left[\,
       \ln\left(m_0^2\over 8T^2\right) - 1  \,\right]
   \,.
\end {equation}

The remaining contribution to the Debye mass is now the quantity
extracted from the perimeter-law fall-off of large adjoint Wilson loops on
the lattice.  Since we want specifically to extract the coefficient of
the $O(g^2 T)$ contribution to the Debye mass, we only need the
leading-order result for the perimeter-law exponent in the limit that
$g$ is small.  Formally, this is
\begin {equation}
   \Delta m \sim g^2 T \,
   \lim_{g\to0} \> \lim_{|C|\to\infty} \left[
       -{1\over g^2|C|} \ln
       \left<\tr\,{\cal P} \exp\Bigl(i\oint_C
                            d\v x\cdot \v A_{\rm adj} \Bigr) \right>
   \right] \,,
\end {equation}
and $|C|$ is the perimeter of the loop in lattice units.
The limit, however,
diverges logarithmically as $g{\to}0$ because of the physics behind
the logarithm in (\ref{c def}) and
(\ref{E estimate}).  To cure the problem, we simply need
to extract this logarithm explicitly and combine it with
the perturbative contribution
$m_{\rm lat}$:
\begin {eqnarray}
   \mdebye &=& m_0 + g^2 T \left\{ \alpha +
     {\ca\over8\pi} \left[ \ln\left(m_0^2\over8g^4 T^2\right) - 1 \right]
   \right\}   +  O(g^3 T) \,,
\\
\noalign {\hbox {where}}
   \alpha &=& \lim_{g\to0} \> \lim_{|C|\to\infty} \left[
       -{1\over g^2|C|} \ln
       \left<\tr\,{\cal P} \exp\Bigl(i\oint_C
                          d\v x\cdot \v A_{\rm adj} \Bigr) \right>
       + {\ca\over8\pi} \ln \left(g^4\right)
   \right] \,.
\end {eqnarray}
This is our final result for the $O(g^2 T)$ contribution to
the Debye mass expressed in terms of the perimeter law coefficient for
large adjoint-representation Wilson loops in three-dimensional pure
lattice gauge theory.
All that is needed to obtain a numerical result is for someone to compute
the value of $\alpha$ on the lattice for gauge theories of
interest [namely SU(3) and SU(2)].


\bigskip

We thank Alexei Abrikosov, Lowell Brown, and David Kaplan
for useful conversations.
This work was supported by the U.S. Department of Energy,
grant DE-FG06-91ER40614.


\appendix
\section*{}

To do the integral in (\ref{lattice integral}), first do the $p_y$
integral, which is straightforward and gives
\begin {equation}
  I(r) \equiv
    \int {d^2p_\perp \over (2\pi)^2} \>
    \Delta_{\rm lat}(p_\perp) \left(1 - e^{i p_x r}\right)
  = {1\over2\pi} \int\nolimits_0^{\pi/2} dq \>
    { 1 - \cos(2rq) \over \sin q \sqrt{1+\sin^2q} } \,,
\end {equation}
where $q \equiv p_x/2$.
Now split the integral into two pieces:
\begin {equation}
   I(r) = {1\over2\pi} \int\nolimits_0^{\pi/2} dq
      \left[ {1\over \sin q \sqrt{1+\sin^2q}} - {1\over q} \right]
           [1-\cos(2rq)]
   + {1\over2\pi} \int\nolimits_0^{\pi/2} {dq\over q} \> [1-\cos(2rq)] \,.
\label{an Ir eqn}
\end{equation}
When $r \to \infty$, the first term can be replaced by
\begin {eqnarray}
   {1\over2\pi} \int\nolimits_0^{\pi/2} dq
      \left[ {1\over \sin q \sqrt{1+\sin^2q}} - {1\over q} \right]
   &=& {1\over2\pi} \left[
      {1\over4} \ln \left( 1 - \sqrt{1-\sin^4q} \over 1 + \sqrt{1-\sin^4q}
        \right)
      -\ln q \right] \Biggl|_{q=0}^{\pi/2}
\nonumber
\\
   &=& {1\over2\pi} \left[ -\ln\pi + {\textstyle{3\over2}}\ln2 \right] \,.
\end {eqnarray}
The second integral in (\ref{an Ir eqn}) is straightforward, and the final
result is
\begin {equation}
   I(r) \sim
   {1\over4\pi} \left[ \, \ln \, (8 r^2) 
		       + 2 \EulerGamma
		 \, \right]
   \qquad
   \hbox{as $r\to\infty$.}
\end {equation}


\end {document}